\documentclass{article}
\usepackage{spconf,amsmath,graphicx}
\usepackage{cite}
\usepackage{amsmath,amssymb,amsfonts}
\usepackage{algorithmic}
\usepackage{graphicx}
\usepackage{booktabs}
\usepackage{multicol}
\usepackage{multirow}
\usepackage{textcomp}
\usepackage{array,xcolor,colortbl}
\usepackage{svg}
\usepackage{hyperref}
\usepackage{xcolor}

\usepackage{enumitem}
\setlist{nosep, leftmargin=14pt}

\usepackage{mwe} 

\title{3D MRI Synthesis with Slice-Based Latent Diffusion Models: Improving Tumor Segmentation Tasks in Data-Scarce Regimes}
\name{Aghiles Kebaili $^1$, Jérôme Lapuyade-Lahorgue $^1$, Pierre Vera $^2$ and Su Ruan $^1$}
\address{$^1$ LITIS UR 4108, University of Rouen-Normandy, Rouen, 76000, France\\
$^2$ CLCC Henri Becquerel, Rouen, 76038, France}
\begin{document}
%
\maketitle
\begin{abstract}
Despite the increasing use of deep learning in medical image segmentation, the limited availability of annotated training data remains a major challenge due to the time-consuming data acquisition and privacy regulations. In the context of segmentation tasks, providing both medical images and their corresponding target masks is essential. However, conventional data augmentation approaches mainly focus on image synthesis. In this study, we propose a novel slice-based latent diffusion architecture designed to address the complexities of volumetric data generation in a slice-by-slice fashion. This approach extends the joint distribution modeling of medical images and their associated masks, allowing a simultaneous generation of both under data-scarce regimes. Our approach mitigates the computational complexity and memory expensiveness typically associated with diffusion models. Furthermore, our architecture can be conditioned by tumor characteristics, including size, shape, and relative position, thereby providing a diverse range of tumor variations. Experiments on a segmentation task using the BRATS2022 confirm the effectiveness of the synthesized volumes and masks for data augmentation. Code is available here : \url{https://github.com/Arksyd96/synthesis-with-slice-based-ldm}
\end{abstract}
\section{Introduction} \label{sec:intro}
Deep learning has witnessed remarkable growth in medical imaging, demonstrating its notable effectiveness segmentation tasks across various imaging modalities, including MRI \cite{lundervold2019overview,zhou2021latent}. However, the ongoing challenge of limited access to annotated medical imaging data is a major challenge, primarily due to the rarity of certain pathologies and rigorous medical privacy regulations, consequently leading to a laborious and time-consuming manual delineation of tumor masks by medical professionals. In this context, data augmentation has emerged as an inseparable part of deep learning, enabling models to overcome the limitations associated with a scarcity of training samples and generalize more effectively the data. However, when dealing with complex medical imaging structures, conventional augmentation techniques such as rotations, cropping or noise injection, may introduce deformations, resulting in deviations from the true data distribution. To address these challenges, advanced deep learning-based data augmentation techniques have been proposed, striving to generate synthetic samples that closely resemble real data while preserving the semantic integrity of the medical images \cite{kebaili2023deep,song2021deep}. These models also offer privacy preservation and data anonymization.

Generative Adversarial Networks (GANs) \cite{goodfellow2014generative} have found widespread applications in medical imaging \cite{wang2023fedmed,kebaili2023deep} and have been advocated in numerous literature reviews for data augmentation due to their ability to generate realistic images \cite{chen2022generative}. However, GANs exhibit certain limitations, including learning instability, convergence issues, and the well-documented problem of mode collapse \cite{mescheder2018training}, where the generator produces a limited range of samples. In contrast, Variational Autoencoders (VAEs) \cite{kingma2013auto} have been proposed as an alternative to GANs, offering a more stable training process and a more efficient inference procedure. However, VAEs are also known to produce blurry images \cite{kebaili2023deep} and are incapable of generating high-resolution images. Recently, diffusion models have emerged as a promising method for image synthesis, offering superior image quality and realism compared to GANs while maintaining a good mode coverage \cite{dhariwal2021diffusion}. This has led to the rise of these models and the development of various alternatives, such as the Latent Diffusion Model (LDM) \cite{rombach2022high}. Although these models provide an attractive solution to the challenge of limited training data, a common issue arises from their high computational cost and demanding memory requirements, making them impractical for 3D medical image synthesis. This holds particularly true for diffusion-based models, which are more resource-intensive, presenting challenges in their integration into clinical routines, especially for real-time tasks like data harmonization or imputation. Beyond this, generative models also require a significant amount of data, limiting their feasibility in medical imaging. 

\begin{figure}[t]
\begin{center}
\includegraphics[width=\columnwidth]{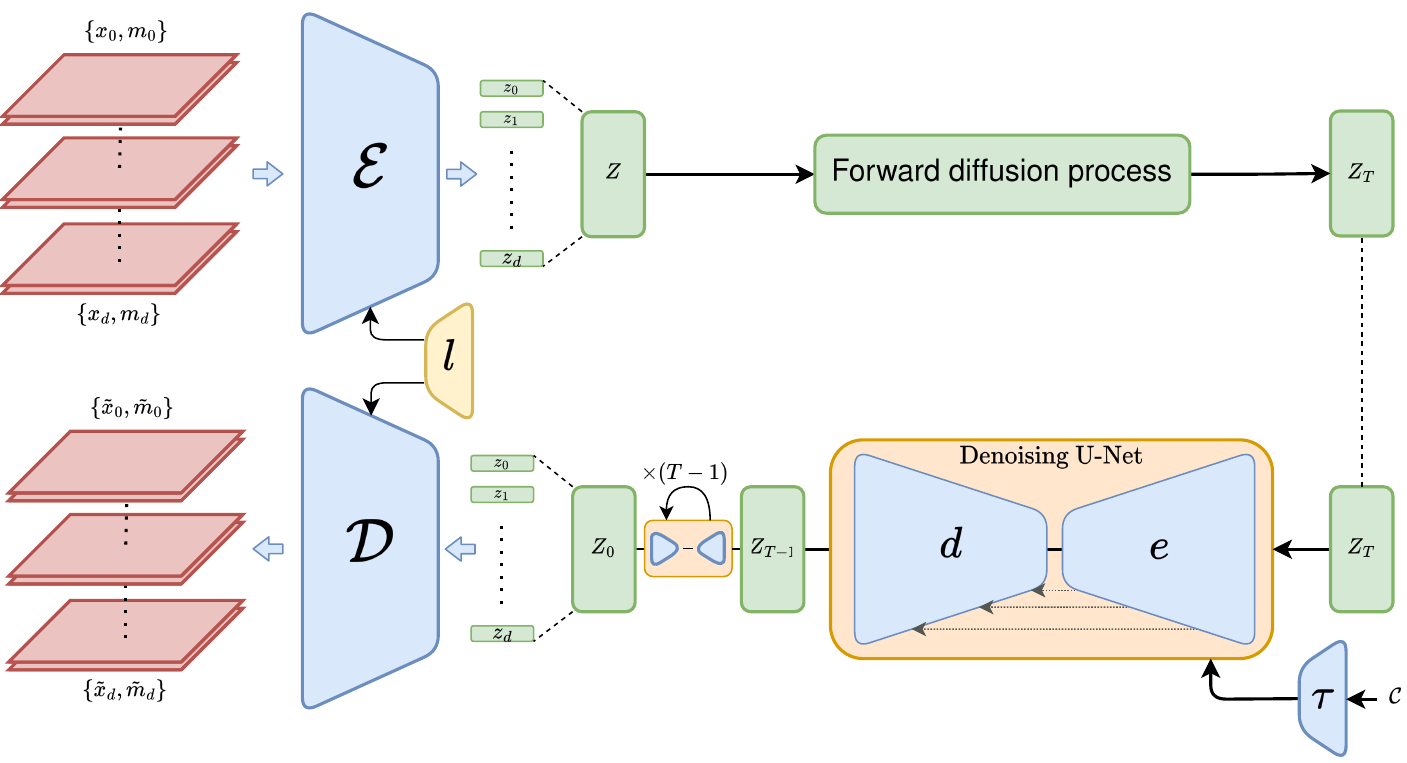}
\caption{Illustration of the proposed architecture. Initially, an MRI volume and its associated 3D mask are decomposed into multiple pairs of 2D slices and masks, denoted as $\{x_i, m_i\}$. These are fed into the encoder $\mathcal{E}$. $l(.)$ and $\tau(.)$ are the positional and condition embedders, respectively. $c$ represents the tumor features vector. See section \ref{sec:sbldm}.} \label{fig:architecture}
\end{center}
\end{figure}


Recent studies have primarily concentrated on image generation or translation \cite{dorjsembe2023conditional,jiang2023cola}, which, in the context of tumor segmentation, is insufficient. The importance lies in generating both images and their corresponding tumor masks, as these masks serve as ground truth for segmentation tasks, adding complexity and cost to the generation process, as we must generalize not only the medical image but also the associated mask. In this study, we introduce a lightweight variant of latent diffusion models (named SBLDM), employing a slice-by-slice approach for the simultaneous generation of medical images and corresponding segmentation masks. Our architecture is trained under data limitations, and we demonstrate its efficacy in augmenting training data for segmentation tasks. Moreover, our model allows precise control of tumor size, shape, and relative position—enabling the generation of a diverse range of tumor variations. This conditionning also serves as regularization to our model. Our evaluation encompasses the quality of generated images, followed by a comprehensive assessment in the context of 3D segmentation tasks using synthesized volumes.

This paper contains three main contributions:
\begin{itemize}
    \item Proposition of an efficient, slice-by-slice diffusion model for the simultaneous generation of high-quanlity medical images and associated segmentation masks with tumor feature controlling.
    \item Highlighting the strength of our approach in data-scarce environments, in contrast to the data-intensive nature of GAN-based architectures.
    \item Comprehensive evaluation showcasing the effectiveness of our high-quality synthesized MRIs in enhancing segmentation tasks.
\end{itemize}

\section{Method} \label{sec:method}
\subsection{Diffusion models}
Diffusion models \cite{sohl2015deep,ho2020denoising} are a subset of generative models based on a forward-and-backward diffusion process. This stochastic process can be thought of as a parameterized Markov chain with a fixed number of time steps, denoted as $T$. During the forward, Gaussian noise is gradually added to an initial data point $x_0 \sim q(x_0)$, following a predefined variance scheduler $\beta_1, \dots, \beta_T$:
\begin{equation}
    q(x_t|x_{t-1}) = \mathcal{N}(x_t; \sqrt{1 - \beta_t} x_{t-1}, \beta_t \mathbf{I})
\end{equation}
During backward, the model is trained to reverse the forward process starting with a Gaussian noise $x_T \sim \mathcal{N}(\mathbf{0}, \mathbf{I})$, and reconstructing it into the initial data distribution $q(x_0)$ with learned parameters $\theta$. This process can be expressed as:
\begin{equation}
    p_\theta(x_{t-1}|x_t) = \mathcal{N}(x_{t-1}; \mu_\theta(x_t, t); \sigma^2_t)
\end{equation}

During training, the model attempts to predict the added noise to $x_0$, denoted as $\epsilon$, and extracts it at each step from $x_t$ to recreate the original sample. 
To learn the parameters $\theta$, such that $p_\theta(x_{t-1}|x_t)$ approximates $q(x_{t-1}|x_t)$, maximum likelihood estimation with variational inference is used, that is similar to VAEs \cite{kingma2013auto} maximizing the evidence lower bound (ELBO). The loss function is finally defined as the mean squared error between the added noise $\epsilon$ and the predicted $\hat{\epsilon}$.

Latent diffusion models (LDM) \cite{rombach2022high} represent a variant of diffusion models that introduces a two-stage process. Initially, data is projected into a lower-dimensional latent space, typically learned through an autoencoder. The diffusion model then operates in this latent space, generating new latent variables. These latents are subsequently transformed back into pixel space using a decoder.

\subsection{Slice-based latent diffusion model (SBLDM)} \label{sec:sbldm}
We propose a new method in response to the difficulties encountered in training 3D diffusion models, arising from the considerable computational expenses and memory constraints, as well as the need for substantial data quantities to avoid overfitting. Our methodology leverages a 2-dimensional VAE with a positional embedder to encode the volumetric data in a slice-by-slice manner. Decomposing volumes into individual slices enables the construction of a larger 2D dataset with increased variance and greater diversity in modes. The accurate positioning of each slice facilitates the use of 2D autoencoders for 3D volume generation, enhancing the autoencoder's capacity to focus on individual slices, and leading to improved generalization.

Our architecture is based on a latent diffusion model that jointly generates 3D MRI volumes and the corresponding tumor mask (Figure \ref{fig:architecture}). A positional embedder $l(.)$ is introduced in our architecture, allowing it to acquire an understanding of the relative position of each slice within the volume. This additional layer of supervision equips the autoencoder with more spatial awareness. The encoder $\mathcal{E}$ is modeled as conditional distribution $q_\phi(z_i | x_i, m_i, l(i))$ where $x_i, m_i \in \mathbb{R}^{W \times H}$ are the image and its corresponding segmentation mask at slice $i \leq D$, and $l(i)$ is the embedding of the slice $i$. The encoder projects samples into a lower-dimensional representation $z_i \in \mathbb{R}^{W^{'} \times H^{'}}$ that encaspulates common characteristics between the image and its associated segmentation mask for a given slice. On the other hand, the decoder $\mathcal{D}$ can be seen as conditional joint distribution $p_\theta(x_i, m_i | z_i, l(i))$ that reconstructs the image and masks pairs given the latent representationsand the relative position of the slice. Subsequently, all these individual $z_i$ are amalgamated to form a 3D latent space, denoted as $Z = \cup_i z_i \in \mathbb{R}^{W^{'} \times H^{'} \times D}$. Subsequently, a diffusion model is  trained to capture not only the broader latent variable distribution but also the implicit volumetric dimension introduced through the concatenation of the latent representations.

\subsection{Conditionning on the tumor characteristics}
We further propose to control our model based on tumor characteristics, allowing us to control the size, shape, and relative position of the tumor. This conditionning also serves as regularization to our model, improving supervision and mitigating overfitting. Additionally, this conditioning helps address scenarios where models might generate tumor-free volumes. By specifying the position, we reinforce the constraint of adding a visible tumor to the synthesized data. The tumor's size and shape are quantified through parameters such as voxel volume, surface area, and sphericity. Meanwhile, its relative position is determined by the coordinates of its center of mass $(x, y, z)$ and its dimensions $(w, h, d)$, collectively forming a bounding box around the tumor. To enable this level of control, we leverage a conditioning vector, which is passed through the Multilayer Perceptron $\tau$ (Figure \ref{fig:architecture}) to encode these parameters into a feature vector. This feature vector is subsequently fused with the main latent representation $Z$ during the diffusion process using a scale-shift norm.

\section{Experimentations} \label{sec:experiments}
\subsection{Dataset}
We evaluate the efficacy of our proposed method using the publicly available dataset: BRAin Tumor Segmentation (BRATS2022) \cite{baid2021rsna} proposes multi-modal MRIs with a volume shape of 240$\times$240$\times$155 and a voxel resolution of 1$\times$1$\times$1 $mm^3$. The images are skull-stripped and co-registered to the same anatomical template. The ground truth segmentation masks are provided for the tumor core (TC), enhancing tumor (ET), and whole tumor (WT) regions forming three tumor labels. In our experiments, we only consider FLAIR modality and the WT region.

\subsection{Implementation details}
We deliberately limited our training set to only 100 volumes to simulate a data-scarce scenario, and evaluations are made on an another set of 100 volumes. To accommodate memory limitations for comparative methods, all volumes were resized to $192 \times 192 \times 96$ dimensions. We employ a VAE as an autoencoder with a downsampling factor of 4 and a U-Net \cite{ronneberger2015u} for the diffusion. Our architecture excludes attention modules and utilizes only one residual block per resolution. The experiments were conducted on an NVIDIA GeForce RTX A6000 GPU with 48GB of VRAM, using the Adam optimizer. We employed a learning rate of $1e-5$. For the segmentation task, we utilize the nnUNet \cite{isensee2018nnu} framework with default settings, including for the standard data augmentation.

\subsection{Quantitative results}
We conducted a quantitative comparison of the generated volumes using our method against two other state-of-the-art techniques. This includes a 3D Least Squares GAN (3D-LSGAN) \cite{xie2016least} with a backbone inspired from Deep Convolutional GAN \cite{radford2015unsupervised} and a 3D version of the original latent diffusion model (3D-LDM) \cite{rombach2022high}, wherein the autoencoder is defined as a VAE-GAN \cite{larsen2015autoencoding} with a downsampling factor of 4. Standard 3D pixel-space diffusion models (DDPM) \cite{ho2020denoising} were excluded due to their memory consumption and unreasonable sampling time. The Structural Similarity Index (SSIM) and Peak Signal-to-Noise Ratio (PSNR), as well as the number of parameters and sampling time of each method are chosen for evaluation. Our results demonstrate that our method achieves the highest SSIM of 0.731 and the top PSNR of 21.701 (see Table \ref{tab:visual_quality}). All accomplished while maintaining an efficient parameter count. Despite its notable efficiency in terms of architecture and sampling time, the 3D-LSGAN experiences the most significant quality impact, primarily due to the data scarcity issue, which makes it impractical for augmentation. GAN-based architectures are notably data-intensive, in contrast to likelihood-based models like 3D-LDM and our approach. To enhance sampling time, we implemented a DDIM sampling scheme \cite{song2020denoising} with our method, limiting the number of steps to 50. This optimization results in notable time savings, with only a negligible loss in quality.

\begin{table}[h]
\centering
\resizebox{\columnwidth}{!}{%
\begin{tabular}{@{}lcccc@{}}
\toprule
Methods & PSNR $\uparrow$ & SSIM $\uparrow$& \#params $\downarrow$ & Sampling time $\downarrow$\\ 
\cmidrule(r){1-1} \cmidrule(l){2-3} \cmidrule(l){4-5}
3D-LSGAN                   & 20.091  & 0.601 & \textbf{71M} & \textbf{0.002s} \\
3D-LDM                     & 21.034  & 0.677 & 728M & 13.324s \\
\cmidrule(r){1-1} \cmidrule(l){2-3} \cmidrule(l){4-5}
SBLDM (Ours)                 & 21.466  & \textbf{0.731} & 159M & 30.900s \\
SBLDM (DDIM)           & \textbf{21.701}  & 0.726 & 159M & 1.965s \\
\bottomrule
\end{tabular}%
}
\caption{Quantitative performance of the proposed generative models on the BRATS datasets. \label{tab:visual_quality}}
\end{table}

\begin{figure}[t]
\begin{center}
\includegraphics[width=\columnwidth]{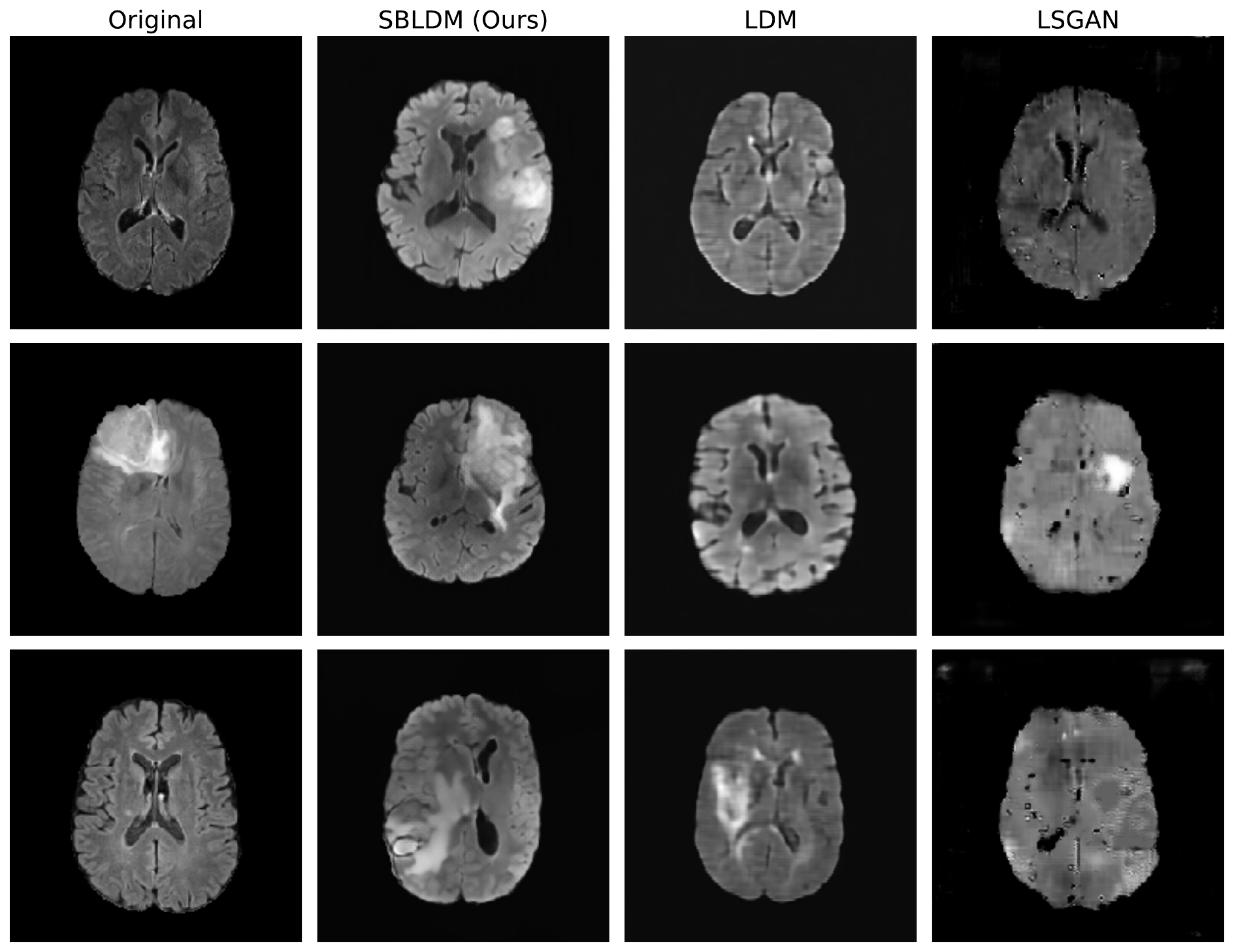}
\caption{Comparison of images generated using our proposed method and other generative models on the BRATS datasets.} \label{fig:comparison}
\end{center}
\end{figure}

\subsection{Qualitative results}
\begin{figure}[t]
\begin{center}
\includegraphics[width=\columnwidth]{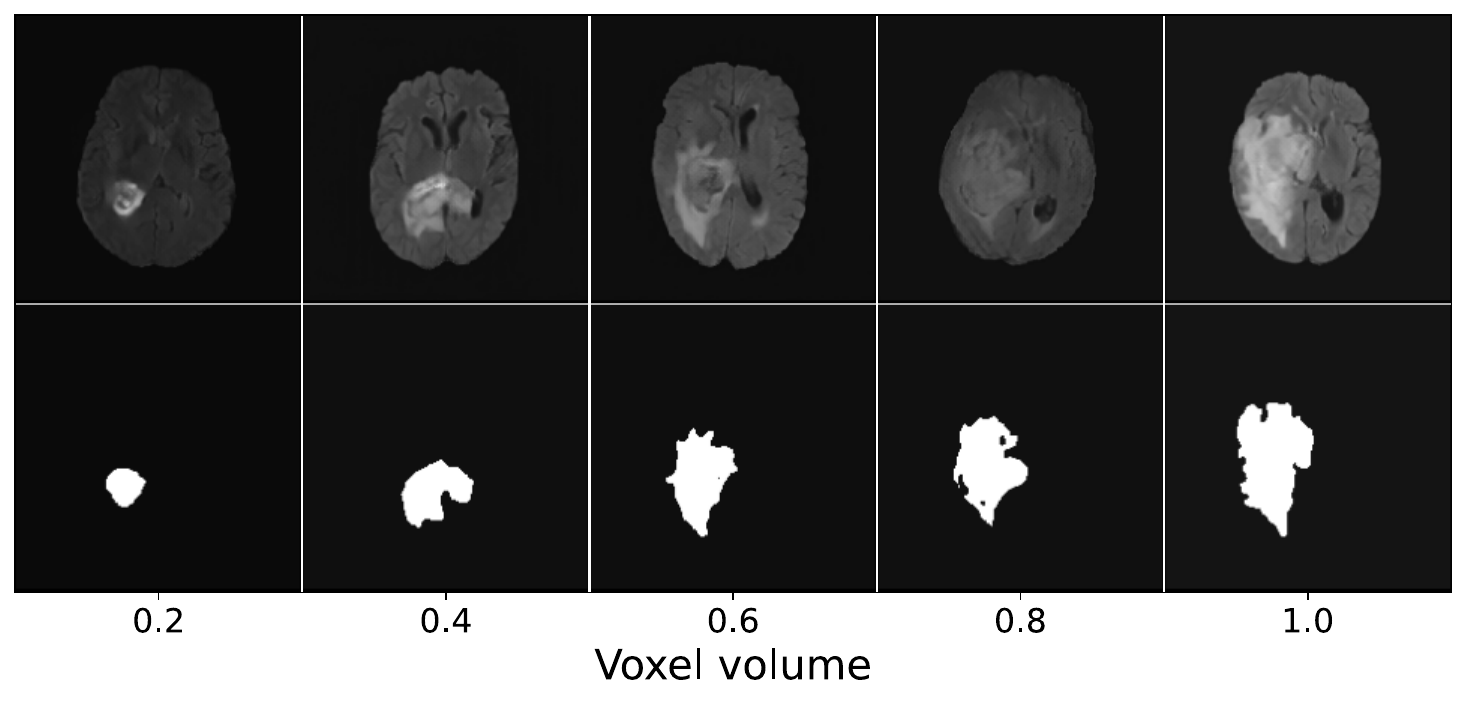}
\caption{Figure showcasing volumes with varying tumor Sizes, ranging from 0.0 to 1.0. The tumor position is fixed.} \label{fig:tumor_sizes}
\end{center}
\end{figure}

We present a qualitative comparison between samples synthesized using different methods from an axial view in Figure \ref{fig:comparison}. Our method's samples exhibit a higher level of realism and fine-grained details, being nearly indistinguishable from real volumes. The mode-coverage is further enriched by the conditionning we offer during the generation process. In contrast, the 3D-LDM's autoencoder trained with only 100 volumes, produces images that are slightly blurrier and lack fine-grained details, indicative of autoencoder underfitting. As for the 3D-LSGAN, the results are subpar, with minimal brain structure and limited details, accompanied by visible image artifacts. Given the scarcity of data, mode collapse is challenging to mitigate at this stage. We do not present images from sagittal and coronal views due to their low quality on the BRATS dataset, rendering them less informative. In Figure \ref{fig:tumor_sizes}, we illustrate some conditioned generations of MRIs and corresponding masks using our method. By varying the voxel volume parameter, we generate tumors ranging from small to large while staying within the same brain region. These results demonstrate not only the brain's variability with each generation but also our method's successful adherence to size and position constraints.

\subsection{Using synthetic volumes on a segmentation task}
We conducted segmentation model training using synthetically generated data, highlighting their potential as effective data augmentors. Our augmentation pipeline involves data generation from the restricted training set and its combination with the synthetically generated images, thereby creating an augmented dataset, as outlined in \cite{kebaili2023deep}.

To evaluate our approach, we initially report results with the original restricted dataset and its augmented version with a factor of $\times$5. Then, we enlarge the real dataset with an additional 100 synthesized volumes, generated through each respective method. We also train the nnUNet using only synthetic volumes to measure their contribution and individual impact. Our findings, presented in Table \ref{tab:seg}, underscore the ability of our synthesized images to improve the segmentation task results, achieving a DSC score of $0.815$ and IoU of $0.715$. This significantly outperforms the corresponding scores for other methods. Notably, the GAN-based architecture appears to deteriorate the results compared to the baseline. This decline can be attributed to the poor quality of the volumes and the occurrence of mode-collapse. Furthermore, we combine both standard augmentation approaches and our synthetic volumes, which leads to even more substantial improvements. We set the number of synthetic volumes to 100 (factor $\times2$), as we observe that beyond this threshold, the improvement in DSC reaches a plateau. This observation can be attributed to the limited number of modes covered by the synthetic images.

\newcolumntype{C}{>{\centering\arraybackslash}p{0.1\textwidth}}
\begin{table}[h]
\centering
\resizebox{\columnwidth}{!}{%
\begin{tabular}{@{}llCC@{}}
\toprule
\multicolumn{2}{c}{Methods} & DSC $\uparrow$ & IoU $\uparrow$ \\ 
\cmidrule(r){1-2} \cmidrule(l){3-4}
\multicolumn{2}{l}{Real volumes}   & 0.714$\pm$0.05  & 0.592$\pm$0.04  \\
\multicolumn{2}{l}{Augmented volumes ($\times$5)} & 0.806$\pm$0.02  & 0.716$\pm$0.02  \\
\cmidrule(r){1-2} \cmidrule(l){3-4}
Synth only  & 3D-LSGAN  & 0.355$\pm$0.13  & 0.302$\pm$0.11 \\
            & 3D-LDM  & 0.529$\pm$0.07  & 0.401$\pm$0.06 \\
            & SBLDM (ours)  & \textbf{0.673$\pm$0.04}  & \textbf{0.551$\pm$0.04} \\
\cmidrule(r){1-2} \cmidrule(l){3-4}
Real + Synth & 3D-LSGAN  & 0.623$\pm$0.08  & 0.514$\pm$0.07 \\
            & 3D-LDM  & 0.705$\pm$0.03  & 0.583$\pm$0.02 \\
            & SBLDM (ours)   & \textbf{0.815$\pm$0.02}  & \textbf{0.715$\pm$0.02} \\
\cmidrule(r){1-2} \cmidrule(l){3-4}
\multicolumn{2}{l}{Real + SBLDM synth + Aug.} & \textbf{0.834$\pm$0.01}  & \textbf{0.739$\pm$0.01} \\
\bottomrule
\end{tabular}%
}
\caption{Quantitative performance of the segmentation task in Dice coefficient (DSC), and Intersection-over-Union (IoU). \label{tab:seg}}
\end{table}

\section{Conclusion}
In this paper, we present a slice-based latent diffusion model for the joint synthesis of 3D MRI volumes and their segmentation masks in data-scarce regimes. Our approach offers practical advantages, requiring fewer computational and memory resources compared to traditional pixel-space and standard latent diffusion models. Our future work will focus on adapting our approach for multi-modal MRI synthesis.

\bibliographystyle{IEEEbib}
\bibliography{main}

\end{document}